\documentclass[twocolumn,showpacs,preprintnumbers,amsmath,amssymb]{revtex4}

\usepackage{graphicx}
\usepackage{dcolumn}
\usepackage{bm}

\begin{document}

\preprint{APS/123-QED}

\title{Inhomogeneous extragalactic magnetic fields \\
and the second knee in the cosmic ray spectrum}

\author{Kumiko Kotera}
\email{kotera@iap.fr} 
\author{Martin Lemoine}
 \email{lemoine@iap.fr}
\affiliation{
Institut d'Astrophysique de Paris\\
UMR7095 - CNRS, Universit\'e Pierre \& Marie Curie,\\
98 bis boulevard Arago\\
F-75014 Paris, France}

\date{\today}

\begin{abstract}
Various experiments indicate the existence of a second knee around energy 
$E=3\times 10^{17}$~eV in the cosmic ray spectrum.  This feature could
be the signature of the end of the galactic component and of the
emergence of the extragalactic one, provided that the latter cuts off
at low energies. Recent analytical calculations have shown that this cut-off
could be a consequence of the existence of extragalactic magnetic fields
(Refs.~\cite{L05,AB05}): low energy protons diffuse on extragalactic
magnetic fields and cannot reach the observer within a given time.  We
study the influence of inhomogeneous magnetic fields on the magnetic
horizon, using a new semi-analytical propagation
code. Our results indicate that, at a fixed value of the volume
averaged magnetic field $\langle B\rangle$, the amplitude of the low energy cut-off
is mainly controled by the strength of magnetic fields in the voids of
the large scale structure distribution. 
\end{abstract}

                             
                             
\maketitle

\section{Introduction}
Recent developments in our undertanding of the origin of cosmic rays
have led to the suggestion that the cosmic ray spectrum might comprise
only two components, one of galactic origin, dominant in the energy
range $E\,\lesssim\,10^{17}\,$eV, and another of extragalactic origin
in the energy range $E\,\gtrsim\,10^{18}\,$eV~\cite{BGG02}. In this
interpretation, the cross-over between the two components marks the
so-called ``second knee'' in the all particle cosmic ray spectrum (see
Ref.~\cite{NW00} for a review on cosmic ray data at high energies),
while the ``first knee'' at $E\,\sim\, 2\times 10^{15}\,$eV would be
associated with a change of propagation regime or the maximal energy
of protons at the source. The third feature in the cosmic ray
spectrum, i.e. the ``ankle'' at $E\,\sim\,10^{19}\,$eV would be a
consequence of pair production losses of ultra-high energy protons
accelerated in sources located at cosmological
distances~\cite{BGG02}. 

On purely phenomenological grounds, this modern view is appealing when
compared to the more traditional interpretation in which the ankle is
associated with the emergence of an extragalactic cosmic ray component
out of a more steeply falling spectrum at energies $E\,\lesssim\,
10^{19}\,$eV.  One of these major advantages certainly is the
``economy'' of sources. It is indeed notoriously difficult to
accelerate particles beyond $E\,\approx\,10^{15}\,$eV in supernovae
remnants~\cite{LC83}. Therefore, if the first knee corresponds to the
maximal energy of protons at the source, the fall-off of the galactic
component at $\,\sim\, 10^{17}\,$eV would naturally be associated with
the maximal energy of the iron component at the source. On the
contrary, if the extragalactic component appears at the ankle, one
needs to postulate the existence of a third cosmic ray component
between $\,\sim\,10^{17}\,$eV and $\,\sim\,10^{19}\,$eV (see for
instance Refs.~\cite{BKMW07} for a recent proposal), or to assume that
supernovae are able to accelerate particles up to the ankle (see for
instance~\cite{BL01}).

The modern interpretation of a transition between the galactic and the
extragalactic component at the second knee does not come without
flaws, however. In particular, the smooth matching of the galactic and
extragalactic components at $E\,\sim\,10^{17}\,$eV bears the
unaesthetic look of fine-tuning. Of course, this problem is generic
to the matching of two distinct components at a point where the slope
steepens; the introduction of a third cosmic ray component would not
help in this respect.  As far as the galactic component is concerned,
the fall-off at $E\,\gtrsim\,10^{17}\,$eV arises as a direct
consequence of the observation of the first knee, as mentioned
above. However one must explain why the extragalactic component
vanishes at energies below the second knee.  In the original scenario
of Berezinsky et al.~\cite{BGG02}, this low energy cut-off
was attributed to physics at the source. In Ref.~\cite{VZ04}, it was
suggested to interpret it as the modulation of the extragalactic flux
due to a galactic magnetized wind, although the
calculations of Ref.~\cite{M05} bring down this cut-off to a too low
energy, $E\,\sim\, 10^{15}-10^{16}\,$eV.

Another possibility, advocated in Refs.~\cite{L05,AB05} is to relate
this cut-off with the influence of extragalactic magnetic fields. As
demonstrated in these studies, if the intensity of extragalactic
magnetic fields is rather modest, say $\langle
B\rangle\,\sim\,10^{-9}\,$G, the diffusion time of particles with
energy $E\,\lesssim\,10^{17}\,$eV from the closest sources (located
at, say $\,\sim\,50-100\,$Mpc) becomes longer than the age of the
Universe. This produces a low energy cut-off in the propagated
spectrum at the required location, which allows to reproduce a smooth
transition at the second knee in agreement with observational data
(see~\cite{L05}).

It has also been argued that if ultra-high energy cosmic rays
comprised a significant fraction of heavy nuclei ($\gtrsim 20\,$\%),
the scenario of a transition at the second knee would loose its merits
as the energy losses would no longer be able to reproduce the ankle
feature~\cite{BGH04,APO05}. However even a solar type (or Galactic
cosmic ray type) chemical composition, with $\sim10\,$\% helium and
only traces of heavier elements, allows to fit the existing data at
the ankle with a single powerlaw spectrum at
injection~\cite{BGG02}. Furthermore, there is no particular reason to expect
the source composition to be enriched in metals.
In any case, future measurements of the
chemical composition will tell~\cite{BGH04,APO07}.

Quantizing the influence of extragalactic magnetic fields on the
spectrum of cosmic rays with energy $E\,\sim\,10^{17}\,$eV is not an
easy task as the propagation times become of the order of a Hubble
time, hence one must account for the effects of expansion. For the
sake of simplicity, Refs.~\cite{L05,AB05} have thus assumed the
magnetic field power to be distributed homogeneously in
space. However, this approximation deserves to be refined since the
magnetic field is most likely distributed as the charged baryonic
plasma. Since the scale of inhomogeneity of large scale structure in
the Universe is comparable to the distance to the closest sources,
$\,\sim\,50-100\,$Mpc, the inhomogeneity of the magnetic field may
affect the conclusions of Refs.~\cite{L05,AB05}. The objective of the
present paper is precisely to address this issue and to study the
scenario put forward in these references in a more realistic
extragalactic magnetic field configuration.
 
This immediately brings forward the difficulty of defining a realistic
distribution of large scale extragalactic magnetic fields, including
the shape and amplitude of a turbulent magnetic cascade. From an
observational point of view, one has been able to measure the strength
of extragalactic magnetic fields ``only'' in the core of clusters of
galaxies~\cite{K94}; a bridge of synchrotron emission on Mpc scales
has been observed in the Coma cluster~\cite{K89}. Hopefully the SKA
project will enlarge considerably the dataset on extragalactic
magnetic fields~\cite{GBF04} but it is not expected to enter operation
before 2017. In the meantime, one thus has to rely on
theory. Unfortunately, the very origin of extragalactic magnetic
fields is unknown, see Ref.~\cite{W02} for a review.  Furthermore,
even if one knew exactly the initial conditions that set the
configuration of magnetic fields at a high redshift, the simulation of
their evolution throughout cosmic history to the present, carrying
sufficient accuracy on a large dynamic range of spatial scales,
remains a formidable task for numerical computing.

In regards of all these uncertainties on the origin of extragalactic
magnetic fields, on their distribution in the present Universe, on the
nature and shape of magnetic turbulence as well as on the transport
properties of particles in chaotic magnetic fields, we adopt a
simplified and parametrized description which allows us to evaluate
the effects of the various sources of uncertainties on the results. As
a by-product of the present study, we thus propose a simple and new
recipe to build semi-realistic magnetic field distributions out of
dark matter simulations (which can be obtained at a lesser cost than
MHD numerical simulations) as well as a new transport scheme which is
more efficient than existing codes in several respects. In particular,
it allows to enlarge artifically the range of scales on which the
magnetic field is distributed, hence to model the influence of
intergalactic magnetized turbulence on particle transport.  These
techniques, which are developed in Section~\ref{propag} and in
Appendix~\ref{num_tech}, allow us to bracket the possible
distributions of extragalactic magnetic fields at the present time and
their impact on the ultra-high energy cosmic ray spectrum. Our results
indicate that, at a fixed value of the volume averaged $\langle
B\rangle$, the amplitude of the low energy cut-off is controlled by
the strength $B_{\rm void}$ of magnetic fields in the voids of the
large structure distribution and the source distance scale $n_{\rm
s}^{-1/3}\,=\,50\,$Mpc. The fact that our conclusions depend more weakly on
other parameters characterizing the magnetic field distribution
provides an adequate a posteriori justification for our semi-analytic
construction. We also argue that this simulation technique offers
various advantages over existing full-blown MHD simulations of large
scale structure formation, at least as far as cosmic ray propagation
is concerned.

The paper is organized as follows. In Section~\ref{propag}, we present
our scheme of inhomogeneous magnetic field simulation and the
numerical technique of cosmic ray transport. We compare these
techniques to existing simulations and discuss the advantages and
drawbacks of each. In Section~\ref{results} we address the issue of
the low energy cut-off in various models of extragalactic magnetic
fields distributions, compute the spectra and compare them to
experimental data. Section~\ref{discussion} discusses the limitations
of our approach and possible future avenues of research. Finally,
Section~\ref{conclusion} summarizes our findings.

\section{\label{propag} Propagation of high energy cosmic rays in extragalactic magnetic fields}

The straightforward way to study the influence of extragalactic
magnetic fields boils down to performing Monte Carlo simulations of
particle propagation in a simulated magnetized Universe. This,
however, brings in two major difficulties, which were alluded to
earlier but which are rarely discussed in the literature: {\it (i)} an
accurate numerical modeling of the transport of charged particles in
magnetic fields; {\it (ii)} an accurate numerical modeling of the
magnetized volume, including magnetized turbulence. 

Point {\it (i)} deals with the theory of cosmic ray diffusion, which in
spite of a long history and recent major progress, has not yet reached
a consensus on the transport of cosmic rays in MHD turbulence (see
Ref.~\cite{LCY02} for a recent review). Actually, the simulation of
particle transport in a well-defined MHD environment is not trivial
even from a purely numerical point of view. For example,
Ref.~\cite{CLP02} has demonstrated that the interpolation of the
magnetic field from a numerical grid gives an erroneous description of
particle transport if the Larmor radius $r_{\rm L}\,\lesssim\, l_{\rm
min}$, where $l_{\rm min}$ represents the grid size, i.e. the minimum
scale of the turbulence inertial range. Point {\it (ii)} deals with
the problem of simulating realistic MHD flows on a large range of
spatial scales, which also constitutes a field of research in its own
right.

\subsection{Magnetic field modeling}

Several pioneering works have studied the propagation of cosmic rays
in so-called ``realistic'' magnetized
environments~\cite{DGST04,DGST05,SME04}. These studies have constructed the
magnetized cube out of a hydrodynamical simulation of large scale
structure formation, which follows the magnetic field in a passive
way for Ref.~\cite{SME04}, and with feedback effects on the matter evolution for Ref.~\cite{DGST05}. The initial conditions for this magnetic field have been set at a
high redshift (although Ref.~\cite{SME04} also models the production
of magnetic fields at accretion shock waves) and the overall amplitude
of this field has been rescaled at the end of the simulation so as to
reproduce the observed strength of magnetic fields in the core of
clusters of galaxies. This ingenious procedure allows to fix the
volume averaged magnetic field independently of the origin of the
magnetic field, although the volume averaged magnetic field now
depends on the details of the amplification process during cluster
formation.

In Ref.~\cite{SME04}, the authors follow the trajectory of cosmic rays
using Monte Carlo methods while the authors of Ref.~\cite{DGST04}
derive an upper bound on the typical cosmic ray deflection using a
semi-analytic transport scheme. Their conclusions are radically
different: the former authors derive a typical deflection of $\,\sim\,
10-20^o$ above $10^{20}\,$eV while the latter find a deflection less
than a degree at these energies. This discrepancy illustrates the
inherent complexity of such simulations. The complexity and the cost
of such numerical simulations are such that it has not been possible
to elucidate the precise origin of this discrepancy yet.  It is likely
that most of this difference is to be attributed to the modeling of
the extragalactic magnetic field, and to a lesser degree, to the
transport scheme.

\begin{figure}
\includegraphics[width=\columnwidth]{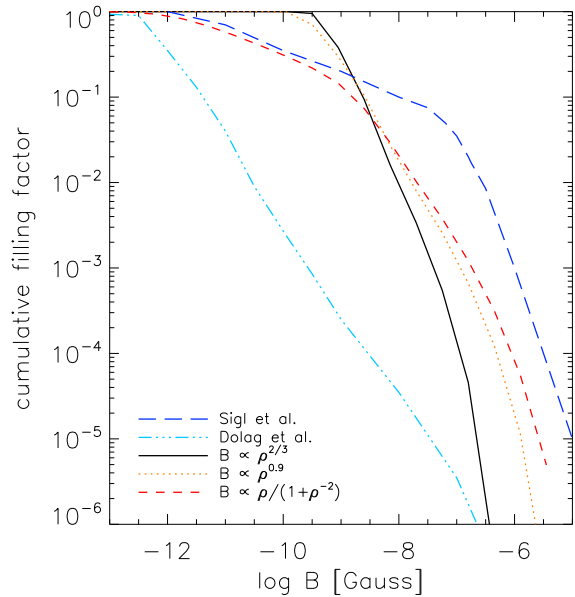}
\caption{\label{fig:Bfill}Volume filling factor of the magnetic field
in different scenarios. In dot-dashed line, the magnetic field
simulated by Dolag and coauthors in Ref.~\cite{DGST04}; in long-dashed line,
that simulated by Sigl and coauthors~\cite{SME04}. In solid line,
 the semi-analytic model with $B\,\propto\,\rho^{2/3}$, in orange dotted line, $B\,\propto\,
\rho$. In red dashed line,
the model $B\,\propto\,\rho\left[ 1 +
(\rho/\overline\rho)^{-2}\right]$; this model simulates a volume with
unmagnetized voids. In all cases the proportionality factor $B_0=2$~nG. }
\end{figure}

  Figure~\ref{fig:Bfill}, which presents the volume filling factors of
the magnetic field strength obtained in these numerical simulations is
particularly instructive (the model of Ref.~\cite{DGST04} is shown as
the dot-dashed line, while the model of Ref.~\cite{SME04} is given
by the long-dashed line). It reveals large differences in the volume
averaged magnetic field as well as in the spatial distribution of
these fields (which translates in this figure as a difference in the
slopes of the volume filling factor). Again, the origin of this
difference is not understood. This figure clearly demonstrates that
{\em the simulated magnetized volumes, despite all the sophistication
of the numerical codes used, cannot truly be deemed as realistic}. It
also indicates the need for alternative methods to study the transport
of high energy cosmic rays in extragalactic magnetic fields, in order
to provide new angles of attack on this difficult problem. This
constitutes one major motivation of the present work, in which we
develop one such  method and apply it to the study of the
low energy cut-off at energies close to the second knee.

Our magnetized volume is constructed in a simple way as compared to
Refs.~\cite{DGST04,SME04}, this simplicity offering various advantages
(and admittedly, several drawbacks) as discussed further below. The
core of our method is to map the magnetic field strength over the gas
density using an analytical relation $B(\rho)$ (to be specified later)
and to distribute randomly the magnetic field orientation in cells of
coherence length $l_{\rm c}$. The gas density itself is obtained from
a high resolution dark matter numerical simulation of large scale
structure formation (with standard cosmological parameters
$\Omega_\Lambda=0.7$, $\Omega_{\rm m}=0.3$ and
$H_0=70\,$km/s/Mpc). Once the volume has been set up, cosmic ray
trajectories are simulated as follows. At each step, the cosmic ray is
supposed to enter a spherical cell of coherence of the magnetic field
defined by its diameter $l_{\rm c}$, in which the magnetic field
orientation is random. The time spent in the cell and the direction of
exit of the cosmic ray are then drawn from semi-analytic distributions
which simulate the transport of the particle in MHD turbulence,
according to studies carried out in Refs.~\cite{CLP02,CR04} (see
Appendix~\ref{num_tech} for a detailed discussion). The particle is
then moved to another coherent cell and the next step is simulated.
As we explain in Section \ref{spectra}, we finally compute the
propagated spectrum in the low energy region
$E\,\lesssim\,10^{17}\,$eV in a semi-analytic way which allows us to
model the effect of cosmological expansion over the course of
propagation from the source to the detector. In detail, we use the
Monte Carlo simulations of particle propagation in the extragalactic
magnetic fields at zero redshift in order to measure the diffusion
coefficients, and use existing analytical formulae in order to
calculate the propagated spectra from the diffusion equation in an
expanding space-time. More details on this latter step are provided in
Section~\ref{spectra}.

Note that the density field of dark matter provides a good
approximation to that of the gas density in the intergalactic medium
(IGM) on scales larger than a few hundred kpc, corresponding to the
baryon Jeans length. Therefore the overall baryonic gas density field
can be obtained by smoothing the dark matter distribution by a window
function of this size. Our dark matter simulation, run with the hydrodynamical code
RAMSES~\cite{T02}, is $512^3$, with extent
$280\,$Mpc, hence with a grid size $\,\simeq\, 560\,$kpc:
in this case the minimum scale of the simulation then plays the role
of the window function and no smoothing is required. Of course, this
treatment does not provide a perfect description of the gas
distribution but we have checked that decreasing the resolution by a
factor 2 does not affect our results. Moreover the gas density field
serves only as a marker of the magnetic field distribution, so that
the above error is negligible in comparison to the uncertainty
surrounding the strength and configuration of the magnetic field. The
essential is rather the law $B(\rho)$ which provides the mapping
between the magnetized volume and the density field. 

In the case of isotropic collapse, it is well known that
$B\,\propto\,\rho^{2/3}$ in a plasma of infinite conductivity. This
law is slightly oversimplistic because it ignores the anisotropy of
collapse in large scale structures, which results in an enhanced
amplification of the magnetic field by shear and anisotropic
compressive flows~\cite{BMT03}. In detail, the equations of ideal MHD
lead to the conservation law:
\begin{equation}
{{\mathrm d}\over{\mathrm d}t}\left({{\mathbf
B}\over\rho}\right)\,=\,\left({{\mathbf B}\over\rho}\cdot{\mathbf
\nabla}\right)\,{\mathbf v}\ ,
\label{eq:MHD}
\end{equation}
which allows to derive the magnification of ${\mathbf B}$ from the
deformation of the density field. If the separation $\mathbf{\delta
q^<}$ between two points is mapped into $\mathbf{\delta
q^>}\,=\,{\cal D}\cdot{\mathbf{\delta q^<}}$ through deformation, where
${\cal D}$ indicates the deformation tensor, then ${\mathbf
B^<}/\rho^<$ is mapped into ${\cal D}\cdot {\mathbf B^<}/\rho^<$, so
that:
\begin{equation}
{B^>\over B^<}\,=\, {\rho^>\over \rho^<}\,{\left\vert {\cal
D}\cdot{\mathbf B^<}\right\vert \over B^<}\ .\label{eq:def}
\end{equation}
From Eq.~(\ref{eq:def}), it is then easy to derive the law $B\,\propto\,
\rho^{2/3}$ for isotropic collapse, or $B\,\propto\, \rho$ for
anisotropic collapse along one (i.e. collapse on a wall) or two
(i.e. collapse on a filament) spatial directions. The law
$B\,\propto\,\rho^{0.9}$ has indeed been observed in the simulations
of Dolag and co-authors~\cite{DGST04}. 

Viscosity and shear flows during collapse may also amplify further the magnetic
field, leading to departures from the law $B\propto \rho$ between
regions of very different density, in particular the voids and the
structures (filaments and pancakes). For instance both simulations of
Refs.~\cite{SME04, DGST04} lead to weaker values of the magnetic fields
in the voids than would be expected from an extrapolation of the law
$B\propto \rho$ to regions of low density. In order to bracket these
different effects, we consider several relations $B(\rho)$:
\begin{eqnarray}\label{eq:Bmodel}
B&\,\propto\,&\rho^{2/3}\ ,\\\label{iso} B&\,\propto\,&\rho^{0.9}\
,\\\label{model2} B&\,\propto\,&\rho\left[1+
\left({\rho\over\langle\rho\rangle}\right)^{-2}\right].\label{model3}
\end{eqnarray} 
The last model is an ad-hoc modeling of the suppression of magnetic
fields in the voids of large structure which leaves unchanged the
distribution in the dense intergalactic medium (meaning
$\rho>\langle\rho\rangle$).

\subsection{Why a semi-analytical propagation method?}

At this stage, one should compare the respective merits and drawbacks
of this new method with other existing techniques. Concerning the
magnetic field distribution, our method obviously neglects subtle
effects such as the amplification of the magnetic field in the
vicinity of accretion shock waves of large scale structure. However,
it should be clear that no numerical simulation can claim to simulate
with accuracy the magnetic field in the vinicity of cosmological shock
waves due to the intricacy of MHD physics at play. The amount of
amplification, the coherence length and the shape of the turbulence
spectrum remain open questions (see however~\cite{SS03} for a
detailed discussion of the Weibel instability operating at
intergalactic shocks).

By considering a one dimensional law $B(\rho)$, our simulation
apparently neglects the influence of the velocity field on the
magnetic field amplification. Indeed, $B$ should be a multidimensional
function that depends on $\rho$ as well as on the velocity field in order to take into account dynamo and shear effects. The three models alluded to earlier do actually account for these effects up to some extent, as they reproduce the characteristic features obtained in the numerical simulations that include dynamo and shear effects. The choice of a random
orientation of the magnetic field in each coherence cell also neglects
the influence of large scale motions. Simulations which follow
explicitely the magnetic field indicate that this latter tends to be
aligned with the principal directions of large scale structure,
i.e. the axis of the filament for example~\cite{B05}. However, one
should note that the simulations of Refs.~\cite{DGST04,SME04,B05}
assume an initial magnetic field with infinite coherence length, so
that the final coherence length of the final magnetic field along the
axis of the filament equals the length of the filament (see Fig.~4 of
Ref.~\cite{B05} for an illustration of this effect). This of
course is unrealistic (unless one assumes an acausal origin for $B$)
since the coherence length of the magnetic field should not exceed
$\,\sim\,1\,$Mpc, the typical turn-around time of an intergalactic
eddy of this size being comparable to the age of the
Universe~\cite{WB99,AB04}.

In contrast, our simulation presents the advantage of simulating this
multiple field reversal along the filament. The alignment of the
magnetic field direction along the filament should take place if the
coherence length is larger than the transverse size of the filament
(more exactly the typical scale height of the density gradient). If,
as is more likely, the coherence length is smaller, then the
compression is similar to planar collapse as far as the magnetic field
in a cell is concerned, hence the field becomes aligned transversely
to the density gradient, and not necessarily along the filament
axis. Our method offers the means to include this effect but we leave
this investigation to future work for simplicity.

A last point concerning the distribution of the magnetic field in the
intergalactic medium is related to its origin. The simulations of
Refs.~\cite{SME04,DGST04,B05} set the initial conditions at high
redshift and ignore other sources of magnetic
pollution of the intergalactic medium (see Ref.~\cite{W02} for a
review on the origin of cosmic magnetic fields), such as galactic
outflows, AGN pollution~\cite{RS68,FL01,GW01}, the amplification of
magnetic fields in accretion shocks of large scale structure (except
Ref.~\cite{SME04} which uses a model of Biermann battery effects),
turbulent amplification in the IGM~\cite{KCOR97}, etc. These sources
should influence the transport of high energy cosmic rays in two ways:
by modifying the relationship between the volume averaged $\langle
B\rangle$ and the value observed in cluster cores, and by adding
additional scattering centers which have been omitted in these
simulations. Our simulation technique offers the freedom to include
such localized pollution effects in a simple and efficient way: one could include these highly magnetized regions in our simulation cube by sampling them according to the local matter density. For
the sake of simplicity, in a first step we ignore these additional
sources and postpone their study to further work. This choice is
conservative in so far as the inclusion of localized regions of
enhanced magnetic field would tend to amplify the magnitude of the low
energy cut-off, all things being equal.

Finally, our simulation assumes for simplicity that the coherence
length $l_{\rm c}$ is uniform in space, whereas it is likely to
evolve as a function of the density and velocity fields. However, this
brings in additional parameters which enlarge the parameter space. We
believe that at this stage, it is more reasonable to study the
influence of $l_{\rm c}$ by performing different runs with different
values of $l_{\rm c}$ and comparing the results. Furthermore, the
actual value $l_{\rm c}$ is intimately related to the origin of the
magnetic field (which sets the initial $l_{\rm c}$) as well as to
the velocity fields which distort the field during the evolution, in
particular with the upbringing of MHD turbulence. Here as well, it
should be clear that no simulation can claim to simulate these various
effects with accuracy. 

The issue of turbulence in the IGM is delicate, because the Reynolds
number in the intergalactic medium may take large or moderate values
depending on the environment. For turbulent excitation on a length
scale $L$ at velocity $v$, this Reynolds number reads~\cite{Lang99}:
\begin{equation}
{\cal R}e\,\simeq\, 10^5\,\left({L\over 1\,{\rm
Mpc}}\right)\left({v\over 300\,{\rm km/s}}\right)\left({T\over
10^5\,{\rm K}}\right)^{-5/2}\left({\rho\over\langle\rho\rangle}\right)\ .
\end{equation}

Turbulence is thus probably fully developed in most of the IGM, except
in the high temperature regions representative of clusters of
galaxies. In these regions of high kinematic viscosity, the shape and
extent of the inertial range of turbulence is rather complex, and most
likely influenced by the strong magnetic field, see Ref.~\cite{SC06}
for detailed discussions. Turbulence plays a fundamental role in the
transport of charged particles as well as in the reshaping of the
distribution of the magnetic field, but its incorporation in numerical
simulations of the gas density is extremely complex.  As the largest
scale of the turbulence cannot exceed a few hundred kiloparsecs or a
megaparsecs, taking into account just one or two decades of inertial
range necessitates an unrealistically high resolution since the cube
size must remain larger than the inhomogeneity scale $\sim
100\,$Mpc. In this respect, our simulations provide more flexibility
because our transport scheme in the simulated magnetic field allows to
simulate the influence of a turbulence spectrum down to scales well
below $r_{\rm L}$, see Appendix~\ref{num_tech}.\\

  The present simulation technique thus combines simplicity,
efficiency with flexibility, and the approximations on which it rests
appear reasonable in regards of the uncertainties surrounding the
origin of extragalactic magnetic fields, the nature of MHD turbulence
and the properties of cosmic ray transport in such turbulence. Most
importantly, its parametrized description allows us to test the
influence of the various parameters on the results in contrast with
most other works on this topic.

\section{\label{results}Results}
The following results were obtained by computing the trajectories of
$10^3$ protons in inhomogeneous magnetic fields mapped according to
four models, for many sets of energies $E$, magnetic field
characteristic values $B_0$ and coherence lengths $l_\mathrm{c}$.  We
will label in what follows ``models 1$-$3'' our modeling of $B(\rho)$
presented in Eqs.~(\ref{iso}$-$\ref{model3}). We add to these models a
last one (model 4) for which $B \propto \rho^{2/3}$ and the level of
turbulence $\eta=\langle\delta{B}^2\rangle/\langle B^2\rangle \ll 1$,
where $\delta{B}$ is the inhomogeneous perturbation component of $B$
(defined such as: $B = \langle B\rangle + \delta{B}$).

Though $B_0$ and $\langle B \rangle$ have quite similar numerical
values, they are not strictly equal (they differ approximately by a
factor 1.5).  $\langle B\rangle$ represents the volume averaged
magnetic field and $B_0$ is the proportionality factor in models
(1$-$4), so that $B = B_0 \times f(\rho)$, where
$f(\rho)$ is dimensionless.

The particles are emitted from $10$ different sources chosen randomly
among regions of high baryonic density. A detailed description of our
code is given in appendix~\ref{num_tech}.

We first spot the existence of a magnetic horizon using the isotropic
collapse magnetic field model (Eq.~\ref{iso}). We then move on to
other models, calculate their resulting propagation spectra and study
the influence of our two main parameters: $B_0$ and $l_{\rm c}$.

\begin{figure}
\includegraphics[width=\columnwidth]{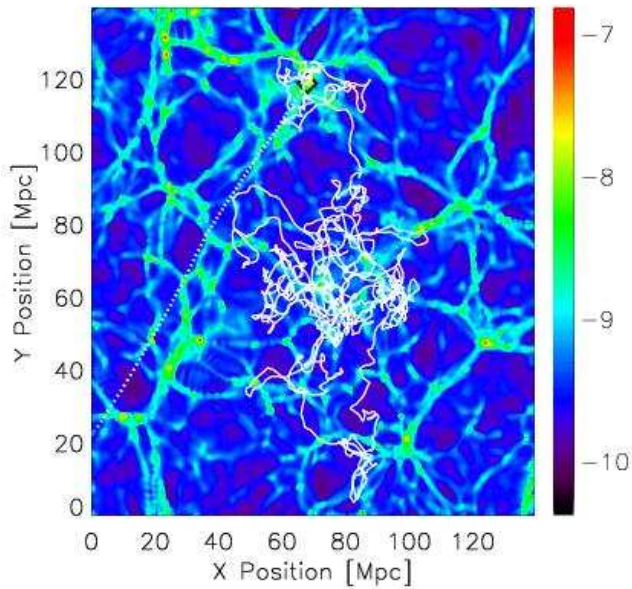}
\caption{\label{slice_propag} Trajectories of protons of different energies
  (solid line: $E = 10^{17}$~eV, dotted line: $E = 10^{19}$~eV) in a slice of
  simulated universe. The characteristic magnetic field
is taken as $B_0=1$~nG and the coherence length as $l_c = 100$~kpc. 
The colorbar on the side indicates the
  intensity of the magnetic field (in log).}
\end{figure}

\subsection{Existence of a magnetic horizon}

Figure~\ref{slice_propag} shows two examples of proton trajectories (solid
line: $E =
10^{17}$~eV, dotted line: $E = 10^{19}$~eV) in a
slice of simulated universe. The characteristic magnetic field
is taken as $B_0=1$~nG and the coherence length as $l_c = 100$~kpc. 

Obviously the particles at $E = 10^{19}$~eV and $E = 10^{17}$~eV evolve
completely differently: the former travel in a rectilinear regime without
being affected by changes in density, whereas the latter experience a
diffusive propagation. Taking a closer look at the diffusive trajectory, one
notices the expected intuitive correlation between the fluffiness of the
trajectory and the clustered regions.\\

\begin{figure}
\includegraphics[width=0.9\columnwidth]{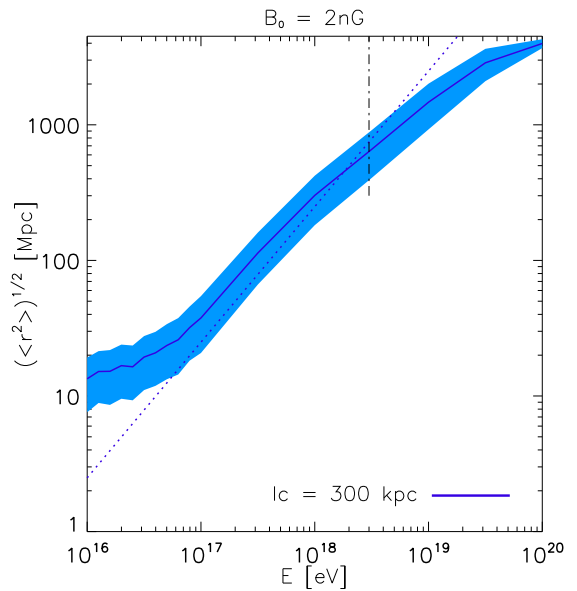}
\caption{\label{travelled_length} Root mean square of the distance of $10^3$
particles to their source after one Hubble time ($t_{\rm H} \sim 13.9$~Gyr) as a function of
their energy, for $B_0=2$~nG and $l_c =
300$~kpc. The solid line represents the root mean
square of the distance and the surrounding color band its variance. The dotted
line shows the values obtained from analytical calculations in a homogeneous
magnetic field (Eq.~\ref{D_semi}) and the dot-dashed line the
threshold energy above which the energy loss time becomes $\lesssim t_{\rm H}/2$. }
\end{figure}

Figure~\ref{travelled_length} illustrates these comments in a more
quantitative way. It shows the root mean square of the distance of $10^3$
particles to their source after one Hubble time (13.9~Gyr) as a function of
their energy, for a characteristic magnetic field $B_0=2$~nG and a coherence
length $l_c =300$~kpc. Our results are no longer valid beyond the dot-dashed
line which represents the threshold energy above which the energy loss time
becomes $\lesssim t_{\rm H}/2$, as our simulations do not compute energy losses.

The first striking remark is that particles of energy below $E \sim
3\times10^{17}$~eV cannot travel farther than a distance of a hundred
megaparsecs from their sources. This corroborates the scenario of
Refs.~\cite{L05,AB05} on the existence of a magnetic horizon, and
extends it to the case of a inhomogeneous magnetic field.

The curve presented in Fig.~\ref{travelled_length} comprises three distinct parts: a
diffusive part with a slope of $\sim 1/6$, a semi-diffusive part (slope $\sim
1$) and a quasi-rectilinear part with a slope tending towards zero. These
trends can be naturally explained by analysing the propagation regimes at
different energies, in {\it homogeneous} magnetic fields. 

The quantity $\langle r^2\rangle^{1/2}$ plotted in Fig.~\ref{travelled_length} can be easily related to the diffusion coefficient
through the equation:
\begin{eqnarray}\label{r_D}
\langle r^2\rangle = 2D t_{\rm H},
\end{eqnarray}
where $t_{\rm H}$ is the Hubble time and $D$ the diffusion coefficient. This
equation follows straightly from the definition of $D\equiv\langle\Delta
x^2\rangle/2\Delta t$, where $\Delta x$ represents the displacement during
the time interval $\Delta t$. 
Thus our computation of $D$ in our simulations has a direct
influence on the shape of the curve observed in Fig.~\ref{travelled_length}.

As explained in appendix~\ref{num_tech}, our diffusion coefficient is
calculated following the results of Casse et al.~\cite{CLP02}. It accounts
for both diffusive ($r_\mathrm{L} \ll l_\mathrm{c}$) and semi-diffusive
($r_{\rm L}>l_{\rm c}$) regimes. In the case of a diffusive regime, equation~(\ref{diff_coeff}) becomes 
\begin{equation}\label{D_diff}
D \propto r_{\rm L}^{1/3}\,l_{\rm c}^{2/3},
\end{equation}
which corresponds to the standard Kolmogorov diffusion regime. Besides, when
the Larmor radius $r_{\rm L}$ is somewhat greater than the coherence length of
the magnetic field, we have the well known dependence
\begin{equation}\label{D_semi}
D \propto r_{\rm L}^2\,l_{\rm c}^{-1}.
\end{equation}
Knowing that $r_{\rm L}\propto E\,B^{-1}$, we get from~(\ref{D_diff}) and
(\ref{D_semi}): $\langle r^2\rangle^{1/2} \propto E^{1/6}$ at low energies and
$\langle r^2\rangle^{1/2} \propto E$ for higher energies. 

It is quite surprising that these slopes, expected for homogeneous magnetic
fields, are also observed in our inhomogeneous simulations for the magnetic
field model and the set of 
parameters presented in Fig.~\ref{travelled_length}. We will show in the
following section that this is not true for other
models and parameters. 

When we get to very high energy ($E\sim 10^{18.5}$), the slope of $\langle
r^2\rangle^{1/2}$ versus $E$ gets weaker, as particles enter the
quasi-rectilinear regime. Eq.~(\ref{r_D}) is no longer valid as
particles never reach the diffusion regime.  \\

\begin{figure}[b]
\includegraphics[width=\columnwidth]{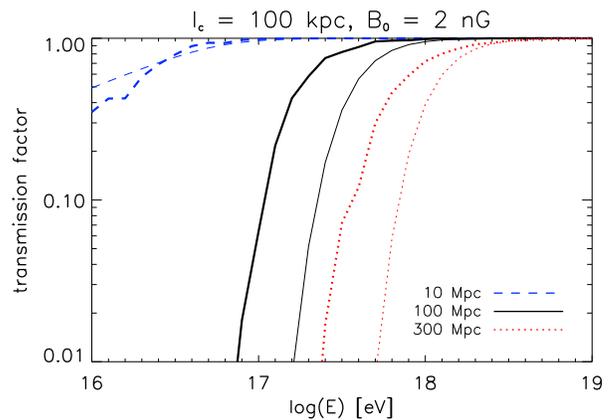}
\caption{\label{transmission} Particle transmission factor at various
  distances from the source, as a function of
  particle energy. Thick lines are results from 
  the simulation run with $B_0 = 2$~nG and $l_\mathrm{c}=100$~kpc. Thin lines
  represent the analytical transmission factor for a homogeneous magnetic
  field (Eq.~\ref{trans}).}
\end{figure}

Another illustration of the existence of the magnetic horizon is presented in
figure~\ref{transmission}. The transmission factor is plotted as a function
of particle energy for three distances to the source (dashed lines: 10~Mpc,
solid lines: 100~Mpc, dotted lines: 300~Mpc). Given an initial source
position, we propagate protons over one Hubble time. At a distance $R$ from
the source, we calculate the transmission factor by taking the ratio between
the number of particles situated beyond $R$ and the total number of particles
that were emitted.  

Figure~\ref{transmission} clearly indicates the presence of a magnetic
horizon: for energies below $\sim 2\times10^{17}$~eV, only half of the emitted
particles reach a distance of 100~Mpc in a Hubble time. The cut-off energy is
lower than for the case represented in Fig.~\ref{travelled_length} due to
the lower value
of $l_{\rm c}$, as will be explained in section~\ref{spectra}. 

Thin lines represent the analytical transmission factors calculated in
appendix~\ref{app_trans} using the diffusion coefficient implemented in our
code (Eq.~\ref{diff_coeff}), for the homogeneous case. 
For a given energy with a particular set
of parameters, one can calculate the corresponding Larmor radius $r_{\rm L}$
and then $D$ using~(\ref{diff_coeff}). It is then easy to obtain $\hat{R}$
and calculate $T$ using~(\ref{trans}). 

For the isotropic collapse model (model 1) and
the represented parameters ($B_0=2$~nG, $l_{\rm c}$ = 100~kpc), there is a
noticeable difference between the homogeneous and the inhomogeneous
cases. The cut-off occurs at lower energy for the inhomogeneous case,
probably due to voids that enable particles to travel farther.

The previous remark does not stand for a travelled distance of 10~Mpc (blue
dashed lines). On the contrary, the transmission factor in an inhomogeneous
magnetic field is lower than in the homogeneous case. This is due to the
influence of the dense environment of the source where particles were
emitted. On a small scale of 10~Mpc, low energy particles have just
escaped the high density region surrounding their source and cannot propagate
as far as in the homogeneous case. The influence of the environment will be
discussed in section~\ref{signatures}.

\subsection{\label{spectra}Calculated spectra}

\begin{figure}
\includegraphics[width=\columnwidth]{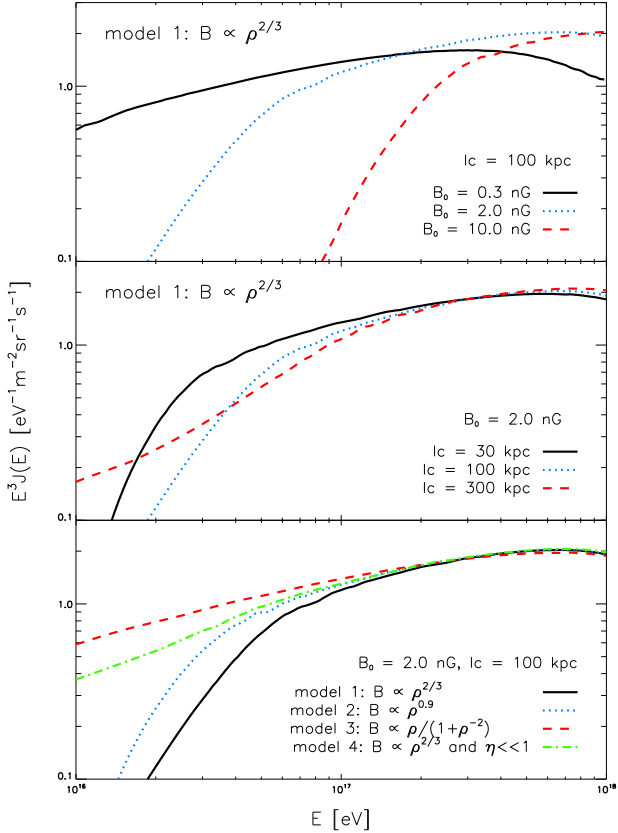}
\caption{\label{spectra_plots} {\it Upper panel:} influence of $B_0$ on the
  spectra for $l_{\rm c} = 100$~kpc and model 1. {\it Middle
  panel:} influence of $l_{\rm c}$ on the spectra for a fixed value of
  $B_0=2$~nG and for model 1. {\it Lower panel:} influence of dependence of
  $B$ over $\rho$ on spectra (models 1$-$4), for fixed values of $B_0=2$~nG and
  $l_{\rm c}=100$~kpc.} 
\end{figure}

\begin{figure}
\includegraphics[width=0.9\columnwidth]{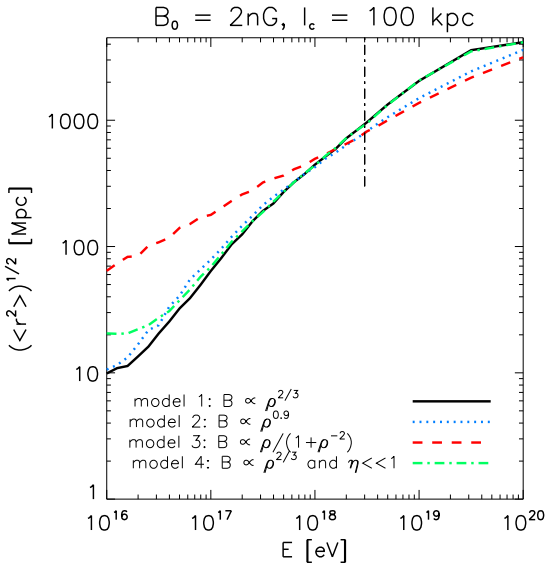}
\caption{\label{travelled_length_many} Same as Fig.~\ref{travelled_length}, for $B_0=2$~nG and $l_{\rm c}=100$~kpc, for models
  1$-$4. Variances are not represented.}
\end{figure}

\begin{figure*}
\includegraphics[width=\textwidth]{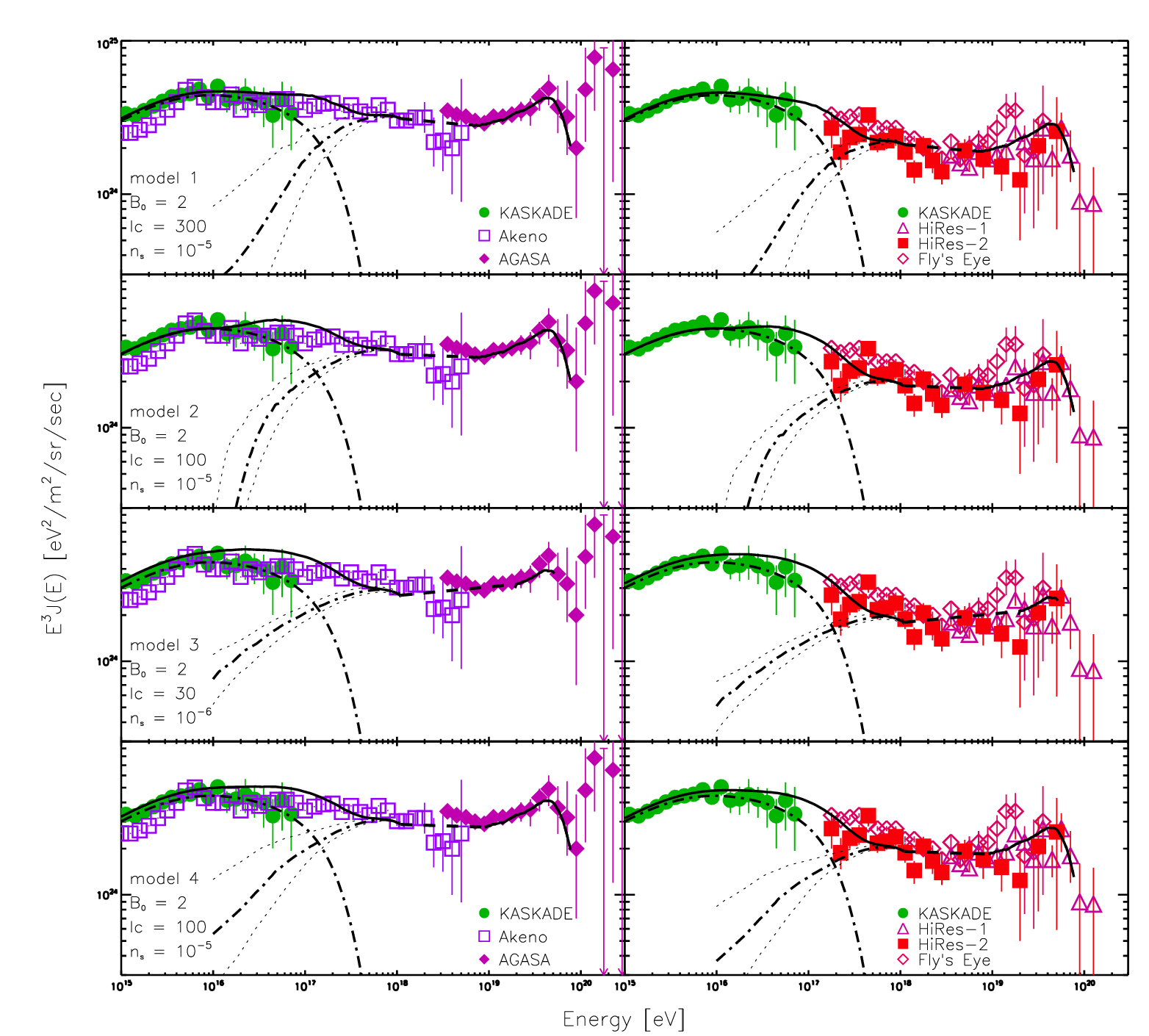}
\caption{\label{many_spectra} Total spectra (galactic + extragalactic)
  compared to data. Each row corresponds to a model and a set of
  parameters. Caution: $n_s=10^{-6}$~Mpc$^{-3}$ for the third row. The left
  panels show KASKADE, Akeno and AGASA data. The right panels show KASKADE,
  HiRes-1, HiRes-2 and Fly's Eyes data. Solid lines represents the median
  values of the total flux, dot-dashed lines the separate galactic and
  extragalactic components and the dotted lines the upper 75$^{\rm th}$ and
  lower 25$^{\rm th}$ percentiles for the magnetic cut-off of the
  extragalactic flux.}
\end{figure*}

In order to compare our results with observational data, we derive spectra
from our simulations in the following way. Ref.~\cite{L05} has shown that the
solution to the diffusion equation in an expanding Universe, assuming a 
constant comoving distance between scattering centers, and limiting itself 
to energy losses by 
expansion (which is correct at energies below $\sim10^{18.3}\,$eV) takes 
the form:
\begin{eqnarray}\label{Jdiff}
J_{\rm diff} =
\frac{c}{4\pi}\int{\rm d}t\,\sum_i
\frac{e^{-r_i/(4\lambda^2)}}{(4\pi\lambda^2)^{3/2}}
\frac{{\rm d}E_{\rm g}(t,E)}{{\rm d}E} Q(E_{\rm g}(t,E)).\nonumber\\
\mbox{ }
\end{eqnarray}
This solution agrees with 
Ref.~\cite{BG06}, which derives the general diffusion equation in an
expanding 
universe and presents the solutions for various energy losses.

In Eq.~(\ref{Jdiff}) above, $r_i$ represents the comoving distance to
source $i$, $E_{\rm g}(t,E)$ the required energy at time $t$ in order
to have an energy $E$ at $t_0$ given the energy losses, $Q(E_{\rm g})$
the emission rate per source at energy $E_{\rm g}$ and $\lambda$ the
comoving ``path length''. $\lambda$ is defined as:
\begin{eqnarray}
\lambda^2=\int_{t_{\rm e}}^{t_0}{{\rm d}t \over a(t)}\,D\left[{a_{\rm
e}E_{\rm e}\over a(t)}\right],
\end{eqnarray}
where $a_{\rm e}$ is the scale factor at emission and $D$ the
 diffusion coefficient. Physically, $\lambda$ represents the typical
 distance travelled by diffusion, accounting for energy losses.

In order to calculate $\lambda$, we first study the dependence of $D$ on $E$,
$B_0$ and $l_{\rm c}$ using our simulations. We find that each set of
parameters corresponds to a different function $D(E,B_0,l_{\rm c})$. We then
parametrize the evolution of the magnetic configuration as done by Berezinsky
\& Gazizov~\cite{BG07}, as:
\begin{eqnarray}
l_{\rm c}(z)=l_{\rm c}(1+z)\quad \mbox{and} \quad B_0(z)=B_0(1+z)^{2-m},
\end{eqnarray}
where $m$ characterizes the MHD amplification of the field. For simplicity,
we set $m$ to 0 in our calculations. This toy model corresponds to a constant comoving distance between scattering centers and ignores magnetic field amplification during structure formation. In this way, we obtain the required
dependence of $D$ over $t$ and thus the function $D[a_{\rm e}\,E_{\rm
    e}/a(t)]$. 

The function ${\rm d}E_{\rm g}(t,E)/{\rm d}E$ is calculated by
integrating the energy losses, following the calculations of
Berezinsky et al.~\cite{BGG02}.  The injection spectrum extends from
$10^{16}$~eV to $E_{\rm max}=10^{20}$~eV. The function $Q(E_{\rm g}) =
K(E_{\rm g}/E_{\rm max})^{-\gamma}$ gives the emission rate per source
at energy $E_{\rm g}$, $K$ being a normalisation factor such that
$\int {\rm d}E\,EQ(E)=L$, with $L$ the total luminosity, which is
assumed to scale as the cosmic star formation rate from Ref.~\cite{SH03}. We
will assume in our calculation a spectral index of
$\gamma = 2.6$. In any case, it should be pointed out that the choice
of the star formation rate has little influence on our spectra, since
the effects of the magnetic horizon dominates those of the star
formation history on the low energy part of the spectrum. \\

At higher energies, when the comoving light cone distance
$r(t)=\int^{t_0}_t {\rm d}t'/a(t')$ becomes smaller than
$\lambda(t,E)$, the propagation is no longer diffusive and enters the
rectilinear regime. In this case, the propagated spectrum is given by:
\begin{eqnarray}
J_{\rm rect}(E)=\frac{c}{4\pi}\sum_i
\frac{1}{4\pi r_i^2}\,
\frac{1}{1+z_i}\,\frac{{\rm d}E_{\rm g}(t_i,E)}{{\rm d}E}\, Q(E_{\rm g}(t_i,E)),\nonumber\\
\mbox{}
\end{eqnarray}
where $t_i$ is related to $r_i$ by $r_i=\int^{t_0}_t{\rm d}t'/a(t')$,
$r_i$ and $z_i$ denoting the comoving distance and redshift of the
$i^{\rm th}$ source. The factor $1/(1+z_i)$ was omitted in
Ref.~\cite{L05}, but it has no influence whatsoever as $z\ll 1$ when
the rectilinear regime is reached.\\

Figure~\ref{spectra_plots} presents the influence of the
$B(\rho)$-models and of parameters $B_0$ and $l_{\rm c}$ on the
magnetic cut-off. Only the diffusive part of the spectra is
represented there and the fall-off of the curve around $\sim
10^{18}$~eV corresponds to the transition between the diffusive and
rectilinear propagation regimes. We assume continuously emitting
sources with density $n_{\rm s}=10^{-5}$~Mpc$^{-3}$
and plot the
median spectrum obtained over 100 realisations of the source
locations. For each realisation, the location of the first hundred
sources were uniformly sampled. For farther sources, the continuous
source approximation is valid and it was used numerically. 

 The upper panel shows the intuitive result that the greater the mean
magnetic field, the steeper the cut-off. Of course this law is not
restricted to model 1 but is also valid for models 2$-$4.

The middle panel shows interesting features that are in agreement with
Eqs.~(\ref{D_diff}) and (\ref{D_semi}). For a fixed value of $B_0$,
for low energies, particles are in the diffusive regime [see
Eq.~(\ref{D_diff})] and $\langle r^2\rangle$ scales with the coherence
length as $l_{\rm c}^{2/3}$. For higher energies particles are in the
semi-diffusive regime [see Eq.~(\ref{D_semi})] and $\langle r^2\rangle
\propto l_{\rm c}^{-1}$. In other words, the spectrum cuts off more
steeply for lower values of $l_{\rm c}$ for low energies and for
greater values of $l_{\rm c}$ for high energies. This can be seen on
figure~\ref{spectra_plots}: for $l_{\rm c} = 300$~kpc, the spectrum
cuts off at high energy but the slope is shallow for low energies,
whereas for $l_{\rm c} = 30$~kpc, the slope is steep at low energies
but the cut-off starts at lower energies.

In view of these trends, one will have to find a good compromise in
order to obtain satisfactory fits to the observational spectra.

The lower panel illustrates the shape of the cut-off for the four
models previously described. Models 3 and 4 present a much shallower
slope compared to models 1 and 2. The almost total absence of magnetic
field in the large scale structure voids for model 3 and the
cancellation of turbulence in model 4 can explain this. We also notice
that model 1 which has a higher magnetic field intensity in the voids
cuts off in a steeper way than for model 2.\\

For a better understanding of the trends seen in the lower panel of
Fig.~\ref{spectra_plots}, we plot in figure~\ref{travelled_length_many} the root mean square of the distance of
$10^3$ particles to their source after one Hubble time as a
function of their energy, as in Fig.~\ref{travelled_length}, for
models 1$-$4. Variances have the same width for all models; we did not
represent them for clarity. As already mentioned, the functions
represented in this figure are closely related to the diffusion
coefficient $D$ [see Eq.~(\ref{r_D})], which is required to calculate
the spectra in the diffusive regime. \\

In figure~\ref{many_spectra} we present the total spectra
(galactic+extragalactic) compared to the data, for our parameter fit
for each model. As for figure~\ref{spectra_plots}, we draw the median
spectrum (dot-dashed line) obtained over 100 realisations of the
source locations. The upper and lower dotted curves show the 75$^{\rm
th}$ and 25$^{\rm th}$ percentiles around this prediction, meaning
that only 25\% of spectra are higher or lower respectively than
indicated by the curves. This uncertainty is related to the location
of the closest sources.  As explained in Ref.~\cite{L05}, we draw a
straight dashed line in the region slightly above $10^{18}$~eV, where
the propagation is neither rectilinear nor diffusive (see
Ref.~\cite{L05} for more details on this transition zone).

The galactic cosmic ray component is modeled as follows. Supernovae are
accepted as standard acceleration sites, yet it is notoriously difficult to
explain acceleration up to maximal energy ~$10^{18}$~eV. Thus it is assumed
that the knee sets the maximal acceleration energy for galactic cosmic rays:
in this conservative model, the spectrum of species $i$ with charge $Z$ takes
the form $j_Z(E)\simeq (E/E_Z)^{-\gamma_i}\exp(-E/E_Z)$, with $\gamma_i \sim
2.4 - 2.7$, a species dependent spectral index, $E_Z\simeq Z\times2\cdot
10^{15}$~eV \cite{KASKADE}. The total galactic component is obtained as the
sum of elemental spectra, each adjusted to KASKADE data as described in
Ref~\cite{L05}. 

We use the data of six major experiments that measured the cosmic ray fluxes
in our regions of interest: KASKADE (2004 data), with an energy range going from $10^{15}$ to
$10^{17}$~eV~\cite{KASKADE}, Akeno from $10^{15}$ to $10^{18.6}$~eV
\cite{Akeno}, AGASA from $10^{18.5}$ to $10^{20.5}$~eV~\cite{AGASA}, HiRes I
and II from $10^{17.3}$ to $10^{20}$~eV~\cite{HiRes} and Fly's Eyes from
$10^{17.3}$ to $10^{20}$~eV~\cite{Fly}. We split these data in two sets in
order to account for the discrepancy between HiRes and
AGASA. This enables us to have two different normalizations for the
extragalactic flux on the left and right panels. The normalization of
KASKADE data remains the same for both sets. \\

Four main points emerge from Fig.~\ref{many_spectra}.  {\it (i)} The second knee feature appears more or less clearly in the four models, but ultimately remains quite robust to model changes.
{\it (ii)} However, again, the influence of the magnetic field intensity in voids is
obvious: even with a source density of $n_{\rm s}=10^{-6}$~Mpc$^{-3}$,
the goodness of fit of model 3 with the observed spectra is only
marginal. This situation is clearly improved in the other models,
especially if we consider the uncertainty on the position of the
closest sources. {\it (iii)} One might also notice that this last
element has a considerable impact on the cut-off energy, much more
than in the case of the homogeneous magnetic field of
Ref.~\cite{L05}. This is due to the presence of the diffusive regime
at the low energy tail. One can indeed observe in Fig.~\ref{travelled_length_many} the flat diffusive locus at low energies
for models 1 and 4. Phenomenologically, one understands that for these
models, a slight change in the closest source distance can influence
greatly the flux of low energy particles. {\it (iv)} Finally,
comparing our plots for AGASA and HiRes data, we conclude that the
fits are better for the latter. The higher slope above the second knee
break point in the HiRes data as well as the gap of data between the
KASCADE and HiRes ranges make the fitting easier. 

One should emphasize, however, that the above fits were obtained by
hand, not by any optimization procedure due to the computing time
required to compute one spectrum. Therefore, the spectra shown above
do not strictly speaking represent the best fit to the
data. Furthermore, one should also exert some caution when comparing
datasets from different experiments. In Fig.~\ref{many_spectra}, we
chose to plot separately the AGASA and HiRes data because of the
well-known discrepancy, but one cannot exclude a discrepancy between
the energy scales of KASCADE and HiRes for instance, which would shift
one dataset with respect to the other. Given all these uncertainties,
the fits shown in Fig.~\ref{many_spectra} appear satisfactory, except
maybe that of model 3 in which the cut-off always appears too mild.

\section{Discussion}\label{discussion}

\subsection{\label{limits}Current limitations}
As already discussed in section~\ref{propag}, our simulations do not
take into account several different features of extragalactic
magnetic fields, both for the sake of simplicity and because they
are in any case poorly understood and poorly constrained. We thus
mentioned that our fields are related to the gas distribution
according to three models (Eqs.~\ref{iso}$-$\ref{model3}), which are
one-dimensional, that no magnetic source is included, and that the
coherence length is assumed to be uniform in space.

Another point that should be underlined is that we propagate our
particles in a static universe, represented by the final output (at
$z=0$) of a cosmological simulation. In other words, the magnetic
fields do not evolve in time during our simulations. The universe
being more dilute at higher redshifts, the effects of inhomogeneous
magnetic fields may be less important.  A way of improving our results
could be to propagate directly particles in an evolving magnetic
field. Such a method would however be very time consuming and as
explained in section~\ref{propag}, subject to too many uncertainties.
One could also apply our semi-analytical propagation method to series
of snapshots of the density of the universe at various $z$.  But
again, one stumbles over our lack of knowledge about extragalactic
magnetic fields: we have no hint of the evolution in time of the
relation $B(\rho)$.  Considering all these uncertainties, our
restriction to a simple static universe thus appears reasonable.

Finally, we did not account for energy losses during the simulations
which serve us to ``measure'' the diffusion coefficient, but included
them in the calculation of the spectra presented in Section~\ref{spectra}.  For energies below $E\sim 10^{18}$~eV that are of
interest to us in this paper, only expansion losses play a noticeable
role (see~\cite{BGG02}), at least at low redshift ($z<1$). Above $\sim
10^{18}$~eV and for greater redshifts, energy losses by photopion and
pair production are no longer negligible. Accounting for these energy
losses should soften the low energy part of Fig.~\ref{travelled_length} if the magnetic field does not evolve strongly
with redshift. Indeed, some of the particles at $E\sim 10^{16-17}$~eV
actually result from higher energy particles that lost their energy.
The greater distance travelled by these particles before losing
their energy would tend to raise the rms of the distance to the source
for low energy particles. Let us stress again that in order to model
this effect, one would need to follow as well the evolution of the
magnetic field with redshift.

\subsection{\label{signatures}Signatures}

\begin{figure}
\includegraphics[width=\columnwidth]{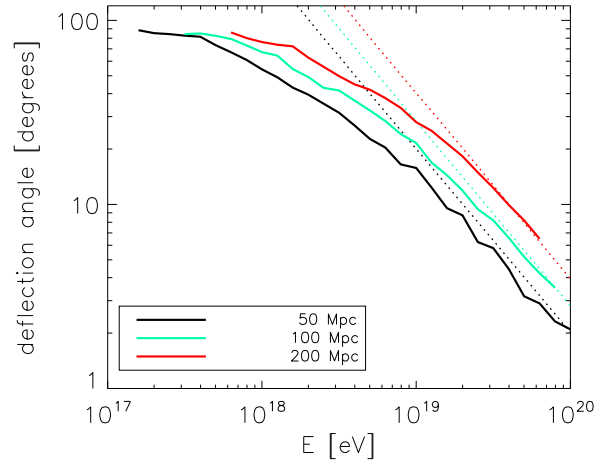}
\caption{\label{deflection} Mean deflection angles at various distances from
  the source, as a function of particle energy, for model 1. Solid lines are the results
  of our simulation with $B_0=2$~nG, $l_{\rm c}=300$~kpc and using model 1
  (Eq.~\ref{iso}). Dashed lines are the analytical values calculated
  in Ref.~\cite{WM96} (Eq.~\ref{deflec}).}
\end{figure}

Firstly, we showed in section~\ref{spectra} that only some particular
types of magnetic fields were able to reproduce the data, in the context of our study. Namely, for
a source density of $n_{\rm s}=10^{-5}$~Mpc$^{-3}$, the voids of large
scale structures should have a certain level of magnetisation, and
$\langle B\rangle$ should be roughly comprised between 0.3 and 10~nG,
2~nG being a overall satisfactory average intensity.  These numbers should be taken cautiously, remembering all the limitations and unknowns that affect these kind of simulations, as stated in the first two sections. \\

We calculated the Faraday rotation measure (RM) for our four magnetic field
models with a characteristic magnetic field of $B_0=2$~nG.
Having sampled $10^4$ lines of sight in our simulation cube, we calculated the
median of the RMs along them. For our models, the power laws of median(RM) versus the distance
are steeper (slope $\sim$~1) than that expected for a homogeneous magnetic
field, for which the integration of RM is equivalent to a simple random walk
(slope~$\sim 1/2$). 

We find that at a cosmological distance of $1$~Gpc, the median of our RMs is
of order $\sim 0.03$~rad/m$^2$ for model~1 and of $\sim 0.1$~rad/m$^2$ for
models~2 and 3. These values are consistent with the current observations of
RMs that predict an upper limit of 5~rad/m$^2$~\cite{K94}. It should be
remarked however that the RMs calculated here are subject to high variations
according to the concentration of matter along the line of sight. Though the
distribution of the RMs is sharply peaked around 0, with most of the RMs in
the narrow interval of $[-0.5,0.5]$~rad/m$^2$, we still find some punctual
cases where the RM can diverge from 20 up to 2000~rad/m$^2$.

The use of a median value of RM enables us to get rid of the
undesirable lines of sight that cross high density clusters and that induce
these divergences. These few lines of sight have a dominant contribution
especially if we calculate the variance or the
root mean square of RM, leading to very high artificial values. 

Note that our rotation measures are again calculated for magnetic fields that
do not evolve in time. Hence our median values can be considered as upper
limits, as far as relatively low density regions are observed. \\

Mean particle deflection angles induced by the magnetic fields of
model 1 are presented in Fig.~\ref{deflection}. At a given distance
from the source, we calculate the deflection angle between the arrival
direction and the line of sight to the source. We stop computing the
angles when the energy loss distance becomes greater than the linear
travelled distance.

Our curves compare quite well to the analytical deflection angles calculated
by Waxman and Miralda-Escude~\cite{WM96} (dashed lines):
\begin{eqnarray}\label{deflec}
\langle \theta \rangle \simeq 0.8^o
\left(\frac{E}{10^{20}\mbox{ eV}}\right)^{-1}\left(\frac{l_\mathrm{c}}{1
  \mbox{ Mpc}}\right)^{1/2}\left(\frac{r}{10
  \mbox{ Mpc}}\right)^{1/2}\nonumber\\
\times\left(\frac{B}{10^{-9}\mbox{ G}}\right),\qquad
\end{eqnarray}
where $r$ is the distance to the source. For all distances, the curves
deviate from the analytical model at low energies, when diffusion becomes
important, and saturate at 90$^o$.

The deflections obtained for cosmological distances at high energy are
quite moderate for model 1. We calculated that it is also the case for
models 2 and 4 (deflections are slightly amplified in model 3). For a particle
energy of $5\times 10^{19}$~eV and a magnetic field of $B_0=2$~nG and $l_{\rm
  c}=300$~kpc, we find that the deflection is of order $\sim 3-5^o$ at
100~Mpc for models 1 and 2, and of $\sim 8^o$ for model 3. These
results are consistent with the observations of doublets and triplets
of events by recent experiments and leave room for doing cosmic ray
astronomy. Our models would thus be in agreement with the detection of
counterparts at energies around the GZK cut-off. \\

Recently, a study related to the present work appeared, claiming that
partial confinement in magnetic fields surrounding the source plays an
important role in the cut-off at low energy~\cite{S07}. This possibility had
been put forward in 
Ref.~\cite{L05}, where it was further shown that the
time of escape from the dense source environment could be non
negligible only in a rather contrived situation, since it requires $B
\gtrsim 1\mu$G$\,(l_{\rm c}/10 \mbox{ kpc})^{1/2}\,(L/100\mbox{
kpc})^{-1}$, with $L$ the characteristic scale of the magnetic field
spread around the source.

We do not find such a strong effect in our present simulations, where
the time of confinement remains of order $\sim 1$~Gyr for a particle of energy
$10^{16}\,$eV, or $\sim 300$~Myr for a particle of energy $10^{17}\,$eV. This
effect obviously depends largely on the source environment, and on the
location of the source. Ref.~\cite{S07} samples the source locations
according to the baryon density and therefore tends to favor high
density (cluster) regions. Since the magnetic field in the simulations
of Ref.~\cite{S07} is already quite strong (see Fig.~\ref{fig:Bfill}),
this explains the magnitude of the effect. As noted in
Ref.~\cite{L05}, the search for counterparts will allow to confirm or
exclude this effect by studying the environment of the sources.

Ref.~\cite{S07} also calculates spectra for inhomogeneous magnetic
fields. However, unlike in our work, Ref.~\cite{S07} does not account
for the evolution of the magnetic field due to expansion during the 
propagation. 

Nevertheless it is interesting that Ref.~\cite{S07} finds the spectrum
to maintain its 'universal' shape in the region of moderate energies,
where the transition between the diffusive and the rectilinear regimes
occurs. This is one region which we cannot probe using the
semi-analytical technique of spectrum reconstruction used in the
present work; the results of Ref.~\cite{S07} justify our
interpolation of the spectrum in this region.

\section{\label{conclusion} Conclusion}

We developed a new method combining an efficient propagation scheme and a
simple recipe to build semi-realistic magnetic field distributions. We map
the magnetic field following the baryon density distribution according to
three different models and propagate particles from cell to cell, taking into
account the inner turbulence of each cell, as well as its global magnetic
intensity. This method is much faster than classical trajectory integrations. 

Under the assumption that the emergence of the extragalactic component occurs
at the second knee, we demonstrated that it was possible give rough limits for some
key parameters ($\langle B\rangle$, $l_{\rm c}$), by studying
their effects on the magnetic horizon. 

For our models assuming isotropic or anisotropic collapse, with or without
turbulence (models~1, 2 and 4 described in sections~\ref{propag} and
\ref{results}), we find that our calculated spectra fit the data
satisfactorily. Numerically, for a source density of $n_{\rm s}=10^{-5}$~cm$^{-3}$ we find that an average magnetic field 
$\langle B \rangle=2$~nG is a reasonable value for the three models cited
above, and coherence lengths of 100~kpc (for models~2 and 4) up to 300~kpc
(model~1) provide a good agreement with the data. These numbers should still be taken cautiously, remembering the limitations discussed throughout this paper. 

We showed that the validity of this scenario depends on other
parameters (relative normalisation of data sets, source density) but 
eventually, the strongest constraint comes from the rate of magnetic
enrichment of the low density intergalactic medium (voids). We saw indeed
that model~3, which simulates a volume with unmagnetized voids has a marginal
goodness of fit with the observed spectra, even with a low source density. 

Ultimately, therefore, the success of this scenario for the transition
between the Galactic and extragalactic cosmic ray components depends on
the very origin of intergalactic magnetic fields, and on whether the
voids of large-scale structures have remained pristine or not.
Interestingly, this question is related to the ongoing debate on the
enrichment of the underdense intergalactic medium in metals, since
galactic winds carrying metals also carry significant magnetic fields.
Detailed studies of the intergalactic medium as well as progress on
extragalactic magnetic fields in the coming decade will shed light on
this issue.

\begin{acknowledgments}
We would like to thank St{\'e}phane Colombi who provided us
with the hydrodynamical simulation outputs, and Christophe Pichon for
valuable discussions. 
\end{acknowledgments}

\appendix

\section{Numerical techniques}\label{num_tech}

\subsection{Setting up the magnetic field}
Our basic assumption is that the magnetic field and the matter density have a
similar spatial distribution at large scales. For the reasons detailed in
section~\ref{propag}, we map the magnetic field following four models
(Eqs.~\ref{iso}$-$\ref{model3}). The parameter $B_0$ mentioned along this
paper refers to the proportionality factor of Eqs.~(\ref{iso}$-$\ref{model3}) and indicates the caracteristic intensity of the
field. 

We calculate our magnetic field by applying these formulae to a three
dimensional dark matter overdensity map (at redshift $z=0$) generated by the
hydrodynamical code RAMSES~\cite{T02}. This cosmological 
simulation was based on a the $\Lambda$CDM model that assumes a flat, low
density universe, with $\Omega_\mathrm{m}=0.3$, $\Omega_\mathrm{\Lambda}=0.7$
and Hubble constant $h \equiv H_0/(100$ km/s/Mpc)$ = 0.7$. It models a
200~Mpc.$h^{-1}$ comoving periodic cube split in $512^3$ cells, where the dark
matter overdensity is computed. We do not resolve structures below Jeans
length, which implies that we can identify the computed dark matter
distribution to a gas distribution. 

While propagating our particles, the magnetic field at a given position is
computed using the overdensity of the nearest grid point.

\subsection{Particle propagation}
For each set of parameters, we propagate $10^3$ protons emitted from
10 different sources. The positions of the sources are chosen as
follows: we first select the grid points in our cube that have an
overdensity $\tilde{\rho} > 10$.  Smoothed over our grid of $512^3$
cells for $200^3$~Mpc$^3$, overdensities of $\sim 10^3$ (minimum
density of galaxies) would indeed correspond to overdensities of $\sim
10$. Thus the selected regions have a good probability of belonging to
massive halos.  We randomly choose ten grid points among this subset
and label them as the initial positions of our particles. The initial
direction of the impulsion of a particle is also drawn randomly.

\subsubsection{Computing the trajectory}
\begin{figure}
\includegraphics[width=0.8\columnwidth]{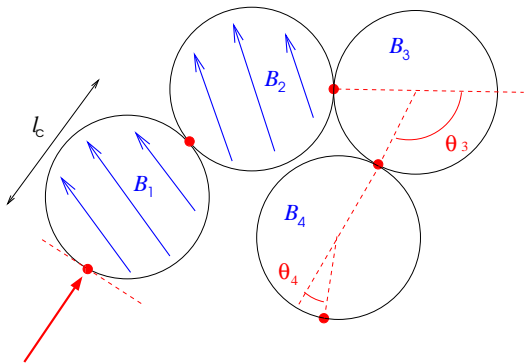}
\caption{\label{bulles} Sketch of particle propagation in magnetic
  fields as modeled in our simulations. Particles enter spheres of
  diameter $l_c$ along a radial direction and escape at calculated
  positions (marked with red dots). A global magnetic field intensity,
  that can be related to the largest scale of turbulence, is
  associated to each sphere (labeled $B_\mathrm{i}$, with
  $B_3>B_4$). Turbulence is taken in account inside each sphere
  through the calculation of the escape time and position. In this
  figure, $\cos\theta_3$ is sampled from a normal law with a large
  variance and $\cos\theta_4$ with a smaller variance. }
\end{figure}

Solving the equation of motion for each particle during an entire
Hubble time using a Runge-Kutta method can be very time consuming. An
alternative method consists in assuming that a proton travels through
adjacent spheres of diameter $l_\mathrm{c}$, in which the magnetic
field has a certain level of turbulence (see Fig.~\ref{bulles}). We
calculate analytically the time of escape from each sphere and sample
the deflection angle of the particle after each sphere from a normal
law where the mean deflection $m$ and its variance $s$ depend on the
Larmor radius of the particle. This sphere-crossing method is much
faster than a direct integration of the trajectory. It also enables us
to take into account the low level turbulence for scales smaller than
$l_\mathrm{c}$.

A particle can cross a sphere in two extreme ways: either $p=
r_\mathrm{L}/l_\mathrm{c} \gg 1$ and the particle goes nearly straight
through the sphere, or $p \ll 1$ and it wanders in the sphere for some
time. In this latter case, the time spent in the sphere $\tau_1
\,\simeq\,l_{\rm c}/c$.

In the Kolmogorov regime in which the diffusion length $l_{\rm
scatt}\,\sim\, r_{\rm L}^{1/3}l_{\rm c}^{2/3}$ with $r_{\rm L}\,\ll\,
l_{\rm c}$, one always have $l_{\rm scatt}\,\leq\,l_{\rm c}$ hence the
particle enters the diffusive regime in the sphere before exiting. The
time spent in the sphere then depends on the diffusion coefficient
$D$.

On average, a particle diffusing through the sphere travels a {\em
linear} distance $l_{\rm c}/\sqrt{2}$, so that the time of escape reads:
\begin{equation}
\tau_2 = \frac{l_\mathrm{c}^2}{4D}\ .
\end{equation}

$D$ is computed according to the results of Casse et al.~\cite{CLP02},
who performed Monte Carlo simulations of particle propagation in
stochastic magnetic fields to measure the spatial diffusion
coefficients. Their data for full turbulence can be fit by the approximate relation:
\begin{eqnarray}\label{diff_coeff}
D = 1.2\,
r_\mathrm{L}c\left(\frac{r_\mathrm{L}}{l_\mathrm{c}}+0.1\left(\frac{r_\mathrm{L}}{l_\mathrm{c}}\right)^{-2/3}\right).
\end{eqnarray}
We use this formula in our code to calculate $\tau$.

\subsubsection{Deflection angle calculation}
Our spheres are located so that the particle always enter
radially. The deflection angle due to the crossing of a sphere is
calculated with respect to this entering direction (see Fig.~\ref{bulles}). 

The cosine of the deflection angle $\cos\theta$ varies according to
the rigidity $p =r_\mathrm{L}/l_\mathrm{c}$. For $p \ll 1$ the
particle is in a diffusive regime, which implies that $\cos \theta$
has a uniform distribution over $[-1,1]$. For the other extreme case,
namely $p \gg 1$, $\cos \theta$ can be sampled from a normal law of
mean $m$ and variance $s$.

In order to identify the functions $m=m(p)$ and $s=s(p)$ in the
quasi-rectilinear regime, we integrated the trajectories of $10^4$
particles in a sphere, using a Runge-Kutta method, for values of $p$
ranging from 2 to 100. This gives us a trend for the mean deflection
angle $\langle \theta \rangle$ and its variance $\delta\theta$.

Then, in our simulations, we use the calculated probability law given
by $m(p)$ and $s(p)$ in order to draw the direction of exit, and we
move the clock forward by the time spent in the sphere.

\subsubsection{Checking the validity of our code in a homogeneous case}

\begin{figure}[t]
\includegraphics[width=\columnwidth]{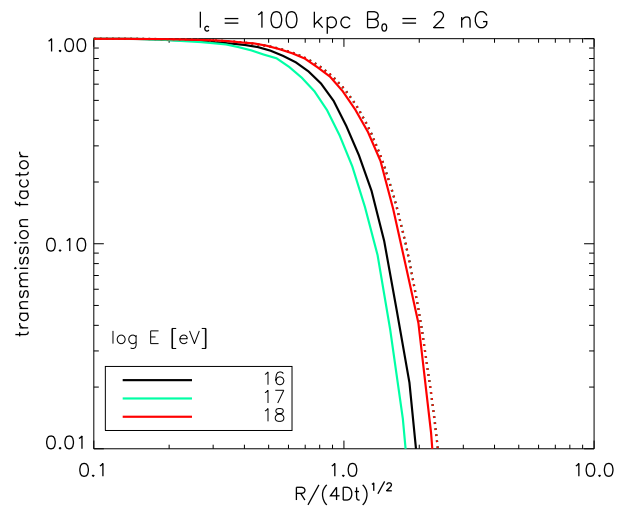}
\caption{\label{Htransmission_vs_Rbar} Fraction of particles located
  at a linear distance greater than $R$ after one Hubble time in a
  homogeneous magnetic field, as a function of $\hat{R}=R/\sqrt{4Dt}$
  as defined in~(\ref{Rhat}), for various rigidities $p=r_{\rm
  L}/l_{\rm c}$. Solid lines are results from our simulations and
  dotted lines the analytical expression calculated in appendix~\ref{app_trans}. }
\end{figure}

Our code was checked on a broad range of parameters in the
analytically calculable homogeneous case, i.e. we compared the
numerical simulations using the above sphere crossing trajectories and
a magnetic field strength equal in each sphere with the analytical
results for full turbulence and homogeneous magnetic power.
We did obtain through the numerical calculations the expected slopes
of 1/6 and 1 for Fig.~\ref{travelled_length} as predicted by the
analytical calculations. The transmission factors also fit nearly
exactly those calculated in Eq.~(\ref{trans}), see Fig.~\ref{Htransmission_vs_Rbar}.

\section{\label{app_trans}Transmission factor for diffusion in homogeneous space}

We calculate in this section the fraction of particles that are
located at a linear distance greater than $R$ at a given time $t$, in
a diffusive propagation regime and in a homogeneous space.  The
density of particles at a position $r$ and at time $t$ can be written:
\begin{eqnarray}
n(r,t) = \frac{1}{(4\pi \,D\,t)^{3/2}}\,\exp\left(-\frac{r^2}{4\,D\,t}\right),
\end{eqnarray}
where $D$ is the diffusion coefficient.  The fraction of particles
beyond $R$ at $t$ can thus be obtained by the following integral:

\begin{eqnarray}
T = \int^{\infty}_{R}4\pi r^2\,{\rm d}r\,\frac{1}{(4\pi
  \,D\,t)^{3/2}}\,\exp\left(-\frac{r^2}{4\,D\,t}\right).
\end{eqnarray}
Integrating by parts, we obtain:
\begin{eqnarray}\label{trans}
T = \sqrt{\frac{2}{\pi}}\,\hat{R}\,\mbox{e}^{-\frac{\hat{R}^2}{2}}+\,\mbox{erfc}\left(\frac{\hat{R}}{\sqrt{2}}\right),
\end{eqnarray}
where 
\begin{eqnarray}\label{Rhat}
\mbox{erfc}(x) = \frac{2}{\sqrt{\pi}}\int^\infty_x\,\mbox{e}^{-u^2}\,{\rm d}u
\quad \mbox{and} \quad \hat{R} \equiv \frac{R}{\sqrt{4Dt}}.
\end{eqnarray}

\newpage
\bibliography{kotera_lemoine}

\end{document}